# Topological charge independent frequency conversion of twisted light

Yan Li[1,2], Zhi-Yuan Zhou[1,2,3,*], Shi-Long Liu[1,2], Shi-Kai Liu[1,2], Chen Yang[1,2], Zhao-Huai Xu[1,2], Yin-Hai Li [1,2,3], Guang-Can Guo[1,2,3], and Bao-Sen Shi[1,2,3,+]

[1]*CAS Key Laboratory of Quantum Information, USTC, Hefei, Anhui 230026, China*

[2]*CAS Center For Excellence in Quantum Information and Quantum Physics, University of Science and Technology of China, Hefei 230026, People's Republic of China*

[3]*Wang Da-Heng Collaborative Innovation Center for Science of Quantum Manipulation & Control, Heilongjiang Province & Harbin University of Science and Technology, Harbin 150080, China*

*\*Corresponding authors: zyzhouphy@ustc.edu.cn; drshi@ustc.edu.cn*

Light with orbital angular momentum (OAM), or twisted light, is widely investigated in the fields of optical communications, quantum information science and nonlinear optics by harnessing its unbounded dimension. For light-matter interacting with twisted light like quantum memory and nonlinear frequency conversion, efficiencies in these processes are usually decreasing exponentially with topological charges, which severely degrades the fidelity of the output states. Here we conceive and develop a method to eliminate the dependence of conversion efficiency on topological charges in second harmonic generation (SHG) process by utilizing a special designed image technique. The independence of SHG conversion efficiency on topological charge is verified for different topological charges, this independence is valid for various pump power. This method can be generalized to other light matter interaction processes and revolute the field of light matter interaction with twisted light to achieve higher efficiency and higher fidelity.

***Introduction*** Orbital angular momentum (OAM) carrying light, or twisted light is one of the most investigated light field since been proposed by L. Allen in 1992 [1]. By harnessing the unbounded dimension and unique mechanical properties of twisted light, various investigations of such kinds of light are performed in many scientific fields [2-4], includes: high capacity spatial mode division multiplexing optical communications [5], high dimensional quantum information processing like quantum memories [6-8] and quantum key distribution [9, 10], nonlinear frequency conversion of twisted light [11-14], high precision optical metrology[15-17], and optical trapping and manipulation of micro-particles [18]. A tremendous researches on different aspects of twisted light are still a very hot topic.

In the field of light matter interaction with twisted light, like quantum memories based on atomic ensembles [6-8] and nonlinear frequency conversion in second [11, 19] or third order nonlinear medium [20], the storage or conversion efficiencies in these processes are strongly depends on the topological charges of the input beam [6, 8, 11, 17], which usually decreasing exponentially with topological charge. There are two main drawbacks for topological charge dependence of conversion efficiency in light matter interactions: on one hand, the decreasing of conversion efficiency with topological charge limits the maximum topological charge that can be efficiently stored or converted; on the other hand, for high dimensional OAM superposition state input, it will severely distorted after storage or frequency conversion. To solve this problem will be a vital important step towards high efficient and high quality light matter interaction with high dimensional superposition states.

Here we conceive and develop a method to eliminate the dependence of conversion efficiency on topological charges in light matter interaction processes by utilizing a special designed image technique. To verify our method, second harmonic generation (SHG) of twisted light in quasi-phase matching periodically poled KTP is shown as an example, but we should point out that the present method can also be generalized to other light matter interaction processes (second or third order nonlinear interaction or quantum storage based on electromagnetic induced transparent) with twisted light. The general theoretical model of our method for SHG is presented first, then we verify the principle by SHG of twisted light with different topological charges by pump the PPKTP crystal with 1560 nm femtosecond fiber laser. Topological charge independent conversion is achieved for topological charges vary from 0 to 3, this is valid for pump power up to 1.1W. The maximum SHG power for different pump modes achieved is 150mW at pump power of 1.1W, which corresponding to conversion efficiency of 14%. The conservation of topological charges are also verified in the conversion processes.

*Theoretical analysis* For SHG of twisted light beam, coupling equation can be used to describe this process [21, 22]. For strong focusing beams or short crystal approximations, the main contribution to second harmonic light is from the pump beam waist, which is determined by

$$E_{2\omega}(x, y, 0) \propto E_{\omega}(x, y, 0)^2 \quad (1)$$

We have assume that phase-matching condition is achieved. Using the relationship between power and electric field

$$P = \frac{n}{2c\mu_0} \int_{-\infty}^{+\infty} \int_{-\infty}^{+\infty} |E|^2 dx dy \quad (2)$$

Where n is the refractive index of the medium *c* is the speed of light at vacuum, $\mu_0$ is the permeability of vacuum. We can obtain

$$P_{2\omega} = \frac{n}{2c\mu_0} \int_{-\infty}^{+\infty} \int_{-\infty}^{+\infty} |E_{2\omega}(x, y, 0)|^2 dx dy \quad (3)$$

By inserting Eq. (1) into Eq. (3), we obtain

$$P_{2\omega} \propto \int_{-\infty}^{+\infty} \int_{-\infty}^{+\infty} |E_{\omega}(x, y, 0)|^4 dx dy \quad (4)$$

If the intensity distribution is independent of topological charge, that is, $|E_{\omega}(x,y,0)|^2$ is independent of topological charge. Then $P_{2\omega}$ is also independent of topological charge. Thus we can achieve conversion efficiency independent of topological charge in SHG. This condition can not be realized for standard modes such as Laggurre-Gaussian (LG) modes. For LG mode the conversion efficiency would decreasing with topological charges [11]. To realize topological charges independent conversion, we need some insight into the generation and evolution of twisted light [23].

Common used methods for generation twisted light are diffractive optical elements like vortex phase plate (VPP) and spatial light modulator (SLM) [3]. These diffractive elements are often illuminates with collimated Gaussian beams. The fact is that when the beam just propagates out from the facet from the diffractive optical element, though it has imprinted with a topologecal charges, it still has an intensity distribution similar to a standard Gaussian beam. This is the key component of our methods. By imaging the field of the light beam just at the output facet of the diffractive elements to the center of the nonlinear crystal with smaller beam size, we can convert the twisted light efficiently without topological charge dependent. The ABCD matrix description of an imaging system should have conditions that *B*=0 and *AD-BC*=1 [24]. In practical their usually two ways for constructing an imaging system, one is 4-f imaging with two lenses, the other single lens imaging. The ABCD matrix for the two systems can be express as

$$\begin{pmatrix} -\frac{f_2}{f_1} & 0 \\ 0 & -\frac{f_1}{f_2} \end{pmatrix} \text{4f imaging;} \quad \begin{pmatrix} -\frac{f}{u-f} & 0 \\ -\frac{1}{f} & 1-\frac{u}{f} \end{pmatrix} \text{single f imaging} \quad (5)$$

Where $f_1$ and $f_2$ are focus length of the two lenses; *f* is focus length of the single lens and *u* is the distance between object and lens. For such imaging systems, the light field in the imaging plane is related to the object plane as [24]

$$E_i(x_i, y_i, z) = \frac{1}{A} \exp(ikz) \exp[\frac{ikC}{A}(x_i^2 + y_i^2)] E_o(\frac{x_i}{A}, \frac{y_i}{A}, 0) \quad (6)$$

Where k is the wave vector of the light field, z is the distance between the imaging plane and the object plane, $x_i$, $y_i$ is the coordinate in the image plane. This formula indicates that despite of a magnification factor A, the intensity distribution of the image plane is the same as object plane. The difference between two lenses imaging and single lens imaging is a phase modulation factor. There is no phase modulation for two lenses imaging. For collimated Gaussian illuminating beam, the light field at the position of VPP can be expressed as [25]

$$E_0(x_o, y_o, 0) = \sqrt{2/\pi} \exp[-(x_o^2 + y_o^2)/w_0^2] \exp[-i\ell Arc\tan(x_o/y_o)] \quad (7)$$

Where $x_o$, $y_o$ is the coordinates at the object plane and $w_0$ is the beam waist for the illuminating Gaussian beam. It is obvious that the intensity distribution at the image plane will be the same for different topological charges by applying Eq. (6), this is the origin for achieving topological charge independent SHG. The beam waist at the image plane would be $Aw_0$. After SHG, the topological charges would be doubled [21]. Below we will verify the above findings.

*Experimental setup* We construct a second harmonic generation experiment from 1560 nm to 780 nm to illustrate the method above. The experimental setup is shown in Fig. 1. A femtosecond fiber laser (Calmer Laser, Mendocino 1560nm) is used as pump laser for SHG process. The central wavelength of pump laser is 1560 nm, the pulse width is 100 fs with repetition rate of 80MHz. A combination of a half wave plate (HWP1) and polarization beam splitter (PBS) function as a power adjustment unit to adjust the pump power for SHG. The maximum average power we can obtain in our setup is about 1.1 W. Another half wave plate (HWP2) is used to change the polarization of pump light to align along the z axes of the nonlinear crystal. A vortex phase plate (VPP) is placed behind HWP2 to imprint different orbital angular momentum on pump beam. Then the pump light is imaged into the center of nonlinear crystal. For simplifying the experimental setup, a single lens (L1) is used for imaging the plane at VPP onto the center of nonlinear crystal. The focal length of L1 is 50 mm and the distance between VPP and L1 is 1 m. The beam waist is about 50 μm in the present configuration. Moreover, this configuration makes the image plane of VPP is almost the same as the focus plane of the lens. So the image plane nearby has the maximum conversion efficiency. We use a type-0 PPKTP crystal for SHG. The crystal has a dimensions of 1 x 2 x1 mm, both end faces of the crystal are anti-reflected coated for both pump and SHG wavelength, the temperature of the crystal is kept at 45.5 $^0$C. Then the pump beam and SHG beam are separated by a dichroic mirror (DM). A collimating lens (L2) transforms the second harmonic light to collimated beam. We use a power meter to measure the power of second harmonic light beam. A camera (BC106-VIS, thorlabs) is used to capture the intensity distribution of SHG beam. To distinguish the topological charge of converted light, a pair of cylindrical lens (CL1, CL2) was used to transform the second harmonic beam to Hermite Gaussian beam. Then we capture the image of transformed light and identify the topological charge of the SHG beam.

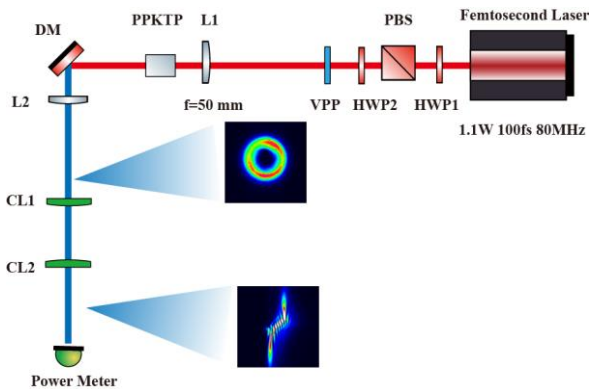

Fig. 1. Experimental setups for the experiment. HWP1, 2: half wave plate; PBS: polarized beam splitter; VPP: vortex phase plate; L1, L2: lenses; DM: dichroic mirror; CL1, 2: cylindrical lenses.

**Results** We obtain conversion efficiency with topological charge range from 0 to 3 at different pump power. Figure. 2 shows the conversion efficiency of different topological charge versus pump power. For a fixed power of pump light the conversion efficiency is almost independent of topological charge. Since the object plane of the output facet VPP1 is imaged into the center of PPKTP crystal, the conversion efficiency of OAM light is almost the same as fundamental Gaussian beam. It illustrates that light with non-Gaussian spatial distribution can have the same conversion efficiency as fundamental Gaussian light. This topological independent phenomenon maintained up to pump power of 1.1 W, the maximum power output of our fiber laser. Conversion efficiency is increasing with pump power, maximum SHG power of 150mW is obtained at 1.1W pump power, which corresponding conversion efficiency of 14%. The small deviation of higher topological charges from the Gaussian beam is arising from slight beam distortion for higher spatial modes because of aberration of single lens imaging. It can be further increased for higher pump power and smaller pump beam waist.

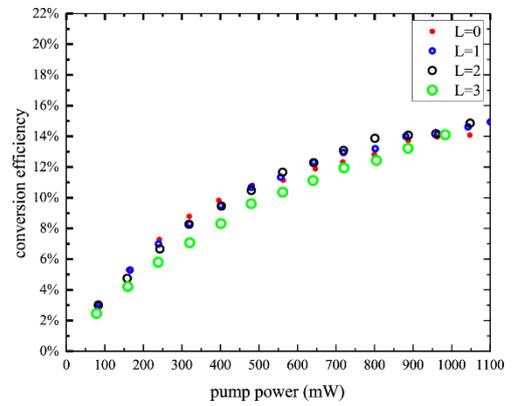

Fig. 2. SHG conversion efficiency as a function of pump power for topological charges ranges from 0 to 3. The small deviation of higher topological charges from the Gaussian beam is arising from slight beam distortion for higher spatial modes because of aberration of single lens imaging.

To verify the topological charge of SHG beam is double of that of pump light, we use a pair of cylindrical lenses to transform the SHG beam. With proper parameters, the transformed light can form an interference-like fringes. The number of dark zones between bright zones is just equal to the topological of SHG beam.

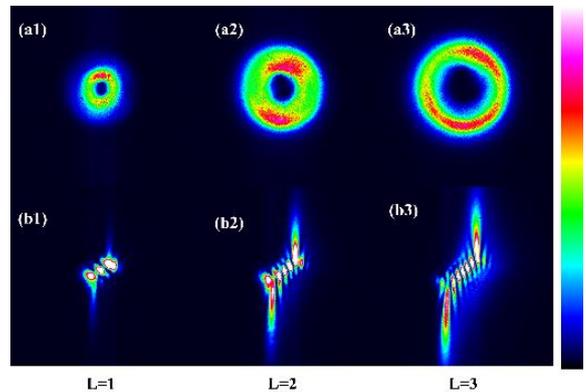

Fig. 3. Images of the SHG beam and the fringes transformed with a pair of cylindrical lenses for pump beam with topological charges of 1, 2, 3.

Fig. 3 shows the images of both SHG beams and its transformed fringes for pump with topological charges of 1, 2, 3. The intensity distribution of second harmonic light likes torus and have larger radius at larger topological charge of pump beam. The fringes indicate that the SHG beams' topological charge are 2, 4, 6, respectively. This confirms that the topological charge of SHG beam is double of that of pump beam and OAM is conserved in SHG.

*Conclusion and discussions* In summary, we have developed a method to eliminate the dependence of conversion efficiency on topological charge in SHG process by imaging the VPP output facet onto the center of nonlinear crystal. We experimentally demonstrate the independence of SHG conversion efficiency on topological charge ranges from 0 to 3. The relationship holds for different pump power up to 1.1W. By identify the topological charges of the SHG beam with cylindrical lens transformation, OAM conservation is verified in SHG.

Further improvement of conversion efficiency can be realized by using high magnification value achromatic objectives to achieving a smaller beam waist or using higher pump power. The present methods can be generalized to other nonlinear processes such as sum frequency generation, different frequency generation, four wave mixing and electromagnetic induced transparency for quantum storage. In the future, we will try to apply this principle to realizing quantum frequency conversion of a high dimensional OAM entanglement states with high efficiency and high quality. The present finds will revolute the field of light matter interaction with twisted light. Which would have significant influence in the field of laser optics and high dimensional quantum information processing.

**Acknowledgments**
This work is supported by the National Natural Science Foundation of China (NSFC) (61605194, 61435011, 61525504); Anhui Initiative In Quantum Information Technologies (AHY020200); the China Postdoctoral Science Foundation (2016M590570).